\documentclass[pra,10pt,twocolumn,superscriptaddress,showpacs]{revtex4-1}
\usepackage{graphicx}
\usepackage{amsmath}
\usepackage{latexsym}
\usepackage{amssymb}
\usepackage{graphicx}
\usepackage[colorlinks=true, citecolor=blue, urlcolor=blue]{hyperref}
\usepackage{float}
\usepackage{amsfonts}
\usepackage{textcomp}
\usepackage{caption}
\usepackage{lipsum}

\newcommand{\ket}[1]{| #1 \rangle}
\newcommand{\bra}[1]{\langle #1 |}

\newcommand{\ketbra}[2]{| #1 \rangle \langle #2 |}

\begin{document}
\title{Can Quantum discord increase in a quantum communication task? }
\author{Shubhayan Sarkar}
\affiliation{School of Physical Sciences, National Institute of Science Education and Research,HBNI, Jatni
              - 752050, India}
            
\author{Chandan Datta} 
\affiliation{Institute of Physics, Sachivalaya Marg,
              Bhubaneswar 751005, Odisha, India.}
\affiliation {Homi Bhabha National Institute, Training School Complex, Anushakti Nagar, Mumbai 400085, India.}
\date{August 13, 2018}

\begin{abstract}
Quantum teleportation of an unknown quantum state is one of the few communication tasks which has no classical counterpart. Usually the aim of teleportation is to send an unknown quantum state to a receiver. But is it possible in some way that the receiver's state has more quantum discord than the sender's state? We look at a scenario where Alice and Bob share a pure quantum state and Alice has an unknown quantum state. She performs joint measurement on her qubits and channel to prepare Bob's qubits in a mixed state which has higher quantum discord than hers. We also observe an interesting feature in this scenario,  when the quantum discord of Alice's qubits increases, then the quantum discord of Bob's prepared qubits decreases. Furthermore, we show that the fidelity of one-qubit quantum teleportation using Bob's prepared qubits as the channel is higher than using Alice's qubits.
\end{abstract}

\maketitle

\section{Introduction}
Non-classical correlations has been identified as a resource for different communication tasks from quantum teleportation \cite{bennett} to remote state preparation \cite{dakic}. Most of the communication tasks uses entanglement as a resource like teleportation, but only a few tasks use quantum discord as a resource. Essentially the use of quantum discord as a resource is not very well explored or is controversial. For instance, let's take the example of remote state preparation. It was shown that separable states can outperform entangled states for remote state preparation. However it was later shown that, its validity is restricted to some special conditions \cite{horodecki}. Quantum discord has also been shown to be generated in the state conversion process even when there is no entanglement \cite{Tame,Koashi}. 

A two qubit mixed state, is called classically correlated \cite{oppenheim} if it can be written as
\begin{equation}\label{classically correlated state}
\rho_{cc}=\sum_{i,j}p_{ij}\ketbra{i}{i}^A\otimes\ketbra{j}{j}^B,
\end{equation} 
where $\ket{i}^A$ are the orthogonal states over the hilbert space $H_A$ and $\ket{j}^B$ are the orthogonal states over the hilbert space $H_B$. However if $\ket{i}^A$ or $\ket{j}^B$ are not orthogonal then the state $AB$ might be non-classically correlated. This led to the idea that mixed states might posses some non-classical correlations which is different from entanglement as the state could be separable. The first non-classical correlations which differ from entanglement is Quantum discord. However, it should be noted that the term ``non-classicality" may have different meanings in quantum information and quantum optics, as was shown in Ref. \cite{Paris,Vogelr} that in the context of quantum optics non-zero quantum discord might have some classical explanations and as well as in Ref. \cite{Sanders} it was shown that classical state may posses some nonzero discord if the measurements are noisy and can be represented by a stochastic channel. In our manuscript the term ``non-classicality" has been used in the language of quantum information where the measurements are not noisy. The notion of Quantum discord was first introduced by Zurek \cite{zurek} and independently by Vedral \cite{vedral} and Horodecki \cite{oppenheim}. Oliver and Zurek then went on to give a measure for Quantum discord \cite{zurek}. Quantum discord was first identified as a resource for computation by Datta {\it et al.}\cite{datta}, where they took a separable state as the resource and showed that the computation was better than using a classical state but not better than using an entangled state. Later Dakic {\it et al.} showed the role of quantum discord in remote state preparation \cite{dakic}, provided that Alice and Bob don't share a reference frame or the operations are bio-stochastic \cite{horodecki}. Many other applications of quantum discord have been proposed whose descriptions can be found in \cite{adesso,streltsov}. 

Olliver and Zurek \cite{zurek} proposed a measure of quantum discord in terms of the mutual information. In classical information theory, one can express the mutual information between two random variables $X$ and $Y$ in two different ways -
\begin{eqnarray}
&&I(X:Y)=H(X)+H(Y)-H(X,Y) \quad \mbox{and}\nonumber\\
&&J(X:Y)=H(X)-H(X|Y),
\end{eqnarray}
where $H(X)=-\sum_xp_x\log_2p_x$ is the Shannon entropy of $X$, $p_x$ is the probability that $X$ takes the value $x$ and $H(X,Y)$ is the joint Shannon entropy of $X$ and $Y$. $H(X|Y)$ is the conditional entropy and defined as $H(X|Y)=\sum_yp_yH(X|y)$, where $p_y$ is the probability of $Y$ taking value $y$ and $H(X|y)$ is the conditional entropy of $X$, such that $Y$ take the value $y$. Unlike classical domain these two expression are different in quantum domain and their difference serve as quantum discord. The generalization of $I(X:Y)$ in quantum theory is 
\begin{equation}
I(\rho^{AB})=S(\rho^A)+S(\rho^B)-S(\rho^{AB}),
\end{equation}
where $I(\rho^{AB})$ is the mutual information between $A$ and $B$ for the state $\rho^{AB}$, $S$ is the von Neumann entropy and $\rho^A$, $\rho^B$ are the reduced density matrix. However, the generalization of $J(X:Y)$ is not that straight forward. Olliver and Zurek \cite{zurek} extended this to quantum realm as 
\begin{equation}
J(\rho^{AB})_{{\Pi_{i}^{B}}}=S(\rho^A)-S(A|{\Pi_{i}^{B}})
\end{equation}
and $S(A|{\Pi_{i}^{B}})$ is given by
\begin{equation}
S(A|{\Pi_{i}^{B}})=\sum_i p_iS(\rho_i),
\end{equation}
where $\rho_i=Tr_B[\Pi_{i}^{B}\rho^{AB}]/p_i$, $p_i=Tr[\Pi_{i}^{B}\rho^{AB}]$ and $\Pi_{i}^{B}$ are the measurement operators on the subsystem $B$. Therefore, the measure of Quantum discord as proposed in \cite{zurek}
\begin{equation}\label{quantum discord}
\delta^{B|A}= \mbox{min}_{{\Pi_{i}^{B}}} [I(\rho^{AB})-J(\rho^{AB})_{{\Pi_{i}^{B}}}],
\end{equation}
where the quantity $I$ represents mutual information and the quantity $J$ represents the amount of information gained about the subsystem $A$ by measuring the subsystem $B$. The minimization occurs over the set of measurement operators such that the quantum discord is measurement independent. For two-qubit states a partial analytic approach was given by Girolami and Adesso \cite{girolami}, however this also needed numerical minimization scheme.

In Quantum teleportation \cite{Liu,Miranowicz} two spatially separated parties Alice and Bob share a quantum channel. Alice has an unknown qubit which she wants to prepare at Bob's end without physically sending it. Then optimizations are done over the measurement basis and channel such that Alice's and Bob's state have greatest degree of overlap (essentially this is maximizing the fidelity). We look at an almost similar scenario where Alice and Bob are spatially separated and have a shared quantum channel. Alice wants to prepare a two qubit state, at Bob's end such that the quantum discord of Bob's state is higher than Alice's state. Moreover, we have shown that when the average quantum discord of Bob's state is higher than Alice's initial state and if we use Bob's state as a resource to teleport a single qubit, instead of Alice's state then the teleportation fidelity increases. As, Quantum discord is very difficult to calculate analytically we used Mathematica (Qdensity~\cite{diaz}) for numerically finding the results.

The paper is organized as follows. Section \ref{sec:sec2} contains the description of the protocol and contains a detailed worked out example with the protocol. A possible application of our result is discussed in section \ref{sec:sec3}  and finally we conclude in section \ref{sec:sec4}. 

\section{Description of the protocol}
\label{sec:sec2}
Alice and Bob share a quantum channel which is a pure state of dimension more than $4$ (for eg. three-qubit $W$ state, or four-qubit cluster state, etc.).
Alice has an unknown quantum state (a two-qubit Werner state $\rho_{A} = \lambda \ket{\phi^+} \bra{\phi^+} + (1-\lambda) I/4$, where $\ket{\phi^+}=\frac{1}{\sqrt{2}}(\ket{00}+\ket{11})$ and $I$ is a four-dimensional identity matrix). Then, Alice jointly measures her two-qubit state and the channel in such a way that Bob receives a two-qubit state. For this joint measurement Alice can choose a basis arbitrarily. As a result she would find different outcomes corresponding to the basis elements. Depending on Alice's outcomes, the two-qubits with Bob will collapse to different states. Alice classically communicates her outcome to Bob. Now Bob measures the quantum discord for each of the separate outcomes and then averages it over all the outcomes. It should be noted here that Bob doesn't need to apply any specific unitary measurement ($U_1\otimes U_2$), as discord remains conserved under a unitary transformation. Also this protocol is different from the remote state preparation in the sense that Alice doesn't know about the state she wants to send to Bob.

We compare the average quantum discord of the states with Alice and Bob and find that average quantum discord of Bob's state is higher than Alice's initial Werner state for most of the range of $\lambda$. We keep the basis chosen by Alice fixed, and vary the parameter $\lambda$ of the Werner state.  

\subsection{Example 1}
\label{sec:sec2.1}
\begin{figure}[t!]
  \includegraphics[width=\linewidth]{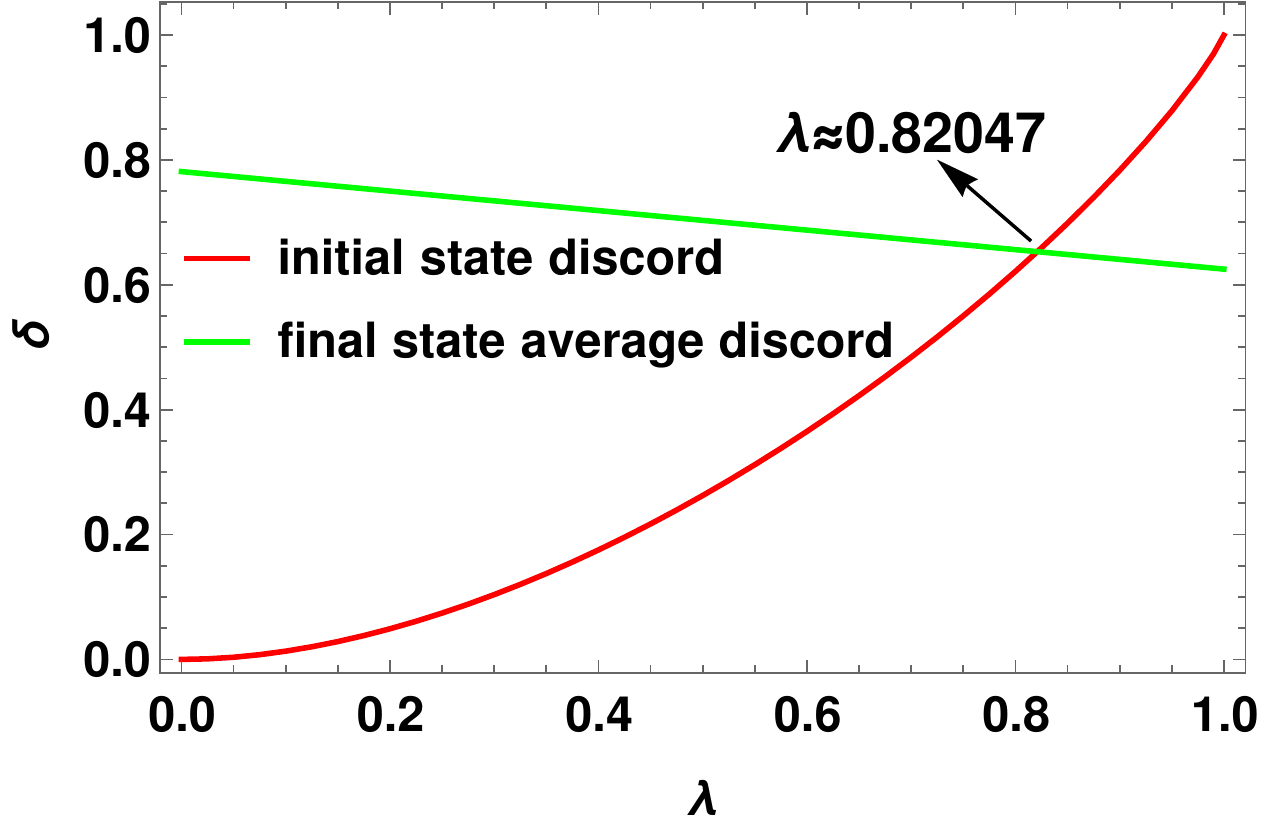}
  \caption{The red line shows the quantum discord as function of parameter $\lambda$ for the initial Alice's Werner state. The green line shows the average discord of the final states with variation in $\lambda$ (using the cluster state (\ref{shared state 1}) as the channel).}
    \label{fig1:}
\end{figure}

Let's take a four-qubit Cluster state 
\begin{equation}\label{shared state 1}
\ket{\Psi}_{C} = 1/2 \Big(\ket{00}_A\ket{00}_B + \ket{01}_A\ket{10}_B + \ket{10}_A\ket{01}_B - \ket{11}_A\ket{11}_B\Big),
\end{equation}
where first two qubits are with Alice and rest two qubits are with Bob. Alice has a two-qubit Werner state  $\rho_{A} = \lambda \ket{\phi^+}\bra{\phi^+} + (1-\lambda) I/4$.
So the total state of the $2^6$ dimensional system is 
\begin{equation}
\rho_{T} = \rho_{A} \otimes \ket{\Psi}_C\bra{\Psi}. 
\end{equation} 
Alice performs joint measurements on the qubits $1,2,3,4$ in the basis which is chosen arbitrarily (we want to project the state of those 4-qubits onto an entangled basis similar to teleportation) and sends 4 bits of classical information to Bob. We are interested in the properties (specifically discord) of the state of the qubits $5,6$ (which is with Bob) after Alice performs her measurement. Alice chooses a complete orthonormal measurement basis as
\begin{eqnarray}\label{four dim basis}
&&\ket{b_1} = \frac{1}{2}\Big(\ket{0001}+\ket{0010}+\ket{0100}+\ket{1000}\Big),\nonumber\\
&&\ket{b_2} = \frac{1}{2}\Big(-\ket{0001}+\ket{0010}+\ket{0100}-\ket{1000}\Big),\nonumber
\end{eqnarray}
\begin{eqnarray}
&&\ket{b_3} = \frac{1}{2}\Big(-\ket{0001}+\ket{0010}-\ket{0100}+\ket{1000}\Big),\nonumber\\
&&\ket{b_4} = \frac{1}{2}\Big(-\ket{0001}-\ket{0010}+\ket{0100}+\ket{1000}\Big),\nonumber\\
&&\ket{b_5} = \frac{1}{2}\Big(\ket{1110}+\ket{1101}+\ket{1011}+\ket{0111}\Big),\nonumber\\
&&\ket{b_6} = \frac{1}{2}\Big(-\ket{1110}+\ket{1101}+\ket{1011}-\ket{0111}\Big),\nonumber\\
&&\ket{b_7} = \frac{1}{2}\Big(-\ket{1110}+\ket{1101}-\ket{1011}+\ket{0111}\Big),\nonumber\\
&&\ket{b_8} = \frac{1}{2}\Big(-\ket{1110}-\ket{1101}+\ket{1011}+\ket{0111}\Big),\nonumber\\
&&\ket{b_9} = \frac{1}{2}\Big(\ket{0000}+\ket{0011}+\ket{1100}+\ket{1111}\Big),\nonumber\\
&&\ket{b_{10}} = \frac{1}{2}\Big(\ket{0000}+\ket{0011}-\ket{1100}-\ket{1111}\Big),\nonumber\\
&&\ket{b_{11}} = \frac{1}{2}\Big(\ket{0000}-\ket{0011}+\ket{1100}-\ket{1111}\Big),\nonumber\\
&&\ket{b_{12}} = \frac{1}{2}\Big(-\ket{0000}+\ket{0011}+\ket{1100}-\ket{1111}\Big),\nonumber\\
&&\ket{b_{13}} = \frac{1}{2}\Big(\ket{0101}+\ket{0110}+\ket{1010}+\ket{1001}\Big),\nonumber\\
&&\ket{b_{14}} = \frac{1}{2}\Big(-\ket{0101}-\ket{0110}+\ket{1010}+\ket{1001}\Big),\nonumber\\
&&\ket{b_{15}} = \frac{1}{2}\Big(\ket{0101}-\ket{0110}-\ket{1010}+\ket{1001}\Big)\,\mbox{and}\nonumber\\
&&\ket{b_{16}} = \frac{1}{2}\Big(\ket{0101}-\ket{0110}+\ket{1010}-\ket{1001}\Big).
\end{eqnarray}
The measurement is mathematically defined as
 \begin{equation}\label{measurement alice}
\hat{M}_{i} = \ketbra{b_i}{b_i} \otimes I,
\end{equation}
where $I$ is the four dimensional identity matrix. Alice sends her outcome to Bob using a classical channel. Then Bob's state would collapse to
\begin{equation}\label{conditional state bob}
 \rho_{B}^i = \mbox{Tr}_{1234}[ \hat{M}_{i} \rho_{T}\hat{M}_ {i}^\dagger],
 \end{equation}
upto some normalization constant $N_i$ which is the probability of occurrence of outcome $\rho^B_i$, is
\begin{equation}\label{normalization constant}
N_i = \mbox{Tr}[\hat{M}_{i} \rho_{T}\hat{M}_{i}^\dagger].
\end{equation}
Then we calculate the quantum discord of the state $  \rho_{B}^i $ as given by \cite{zurek}. Since there are sixteen different outcomes possible for Alice, therefore there are sixteen different $  \rho_{B}^i$'s. Note that as Alice is communicating her results to Bob, Bob's state would collapse to one of the $\rho_B^i$'s. Thus, we compute the discord of each such state and then average it over (which means multiplying the probability $(N_i)$ of the occurrence of state  to the quantum discord of the state $\rho_{B}^i$) to find the average discord $\overline{\delta}$.
\begin{equation}\label{average discord}
\overline{\delta} = \frac{\sum_{i} N_i \delta(\rho_{B}^i)}{\sum_{i} N_i}.
\end{equation}
Interestingly, from the Fig. \ref{fig1:}, we find that upto $\lambda\approx 0.82047$ the average discord of Bob's state is more than the initial Werner state posses by Alice.

\begin{figure}
  \includegraphics[width=\linewidth]{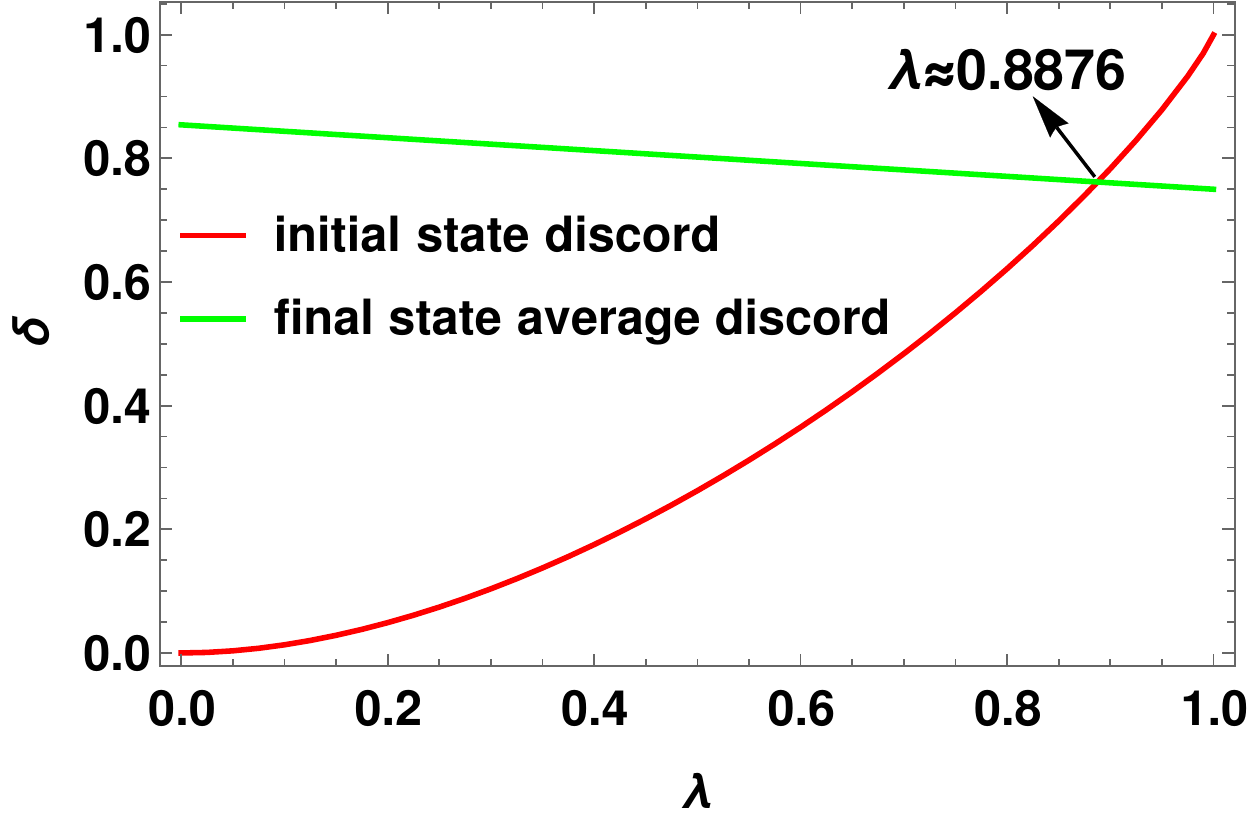}
    \caption{The red line shows the quantum discord as function of parameter $\lambda$ for the initial Alice's Werner state. The green line shows the average discord of the final states with variation in $\lambda$ (using the state (\ref{shared state 2}) as the channel).}
    \label{fig3:}
\end{figure}
Note that for some of the post-selected outcomes $\rho_B^i$'s have a higher Quantum discord than the initial Werner state. Hence, we calculate the average discord and it comes out to be more than that of Alice's initial state for most of the range of $\lambda$. However, here we are not interested in the individual states which have a higher quantum discord
and focus on the quantum discord that Bob would find after averaging over all the outcomes.

\subsection{Example 2}
\label{sec:sec2.2}

We took another 4-qubit channel, 
\begin{eqnarray}
\ket{\Psi}_{\Omega} =&& \frac{1}{\sqrt{6}} \Big(\ket{00}_A\ket{11}_B + \ket{01}_A\ket{01}_B + \ket{01}_A\ket{10}_B+\ket{10}_A\ket{01}_B \nonumber\\&&\quad\quad+ \ket{10}_A\ket{10}_B + \ket{11}_A\ket{00}_B\Big).\label{shared state 2}
\end{eqnarray}
We choose the measurement basis same as in Eq. (\ref{four dim basis}). Here also we get a similar kind of result. From Fig. \ref{fig3:}, it is clear that upto $\lambda\approx0.8876$ the average discord of final state is larger than the initial state.  
  
\subsection{Example 3}  
\label{sec:sec2.3}

\begin{figure}[t!]
  \includegraphics[width=\linewidth]{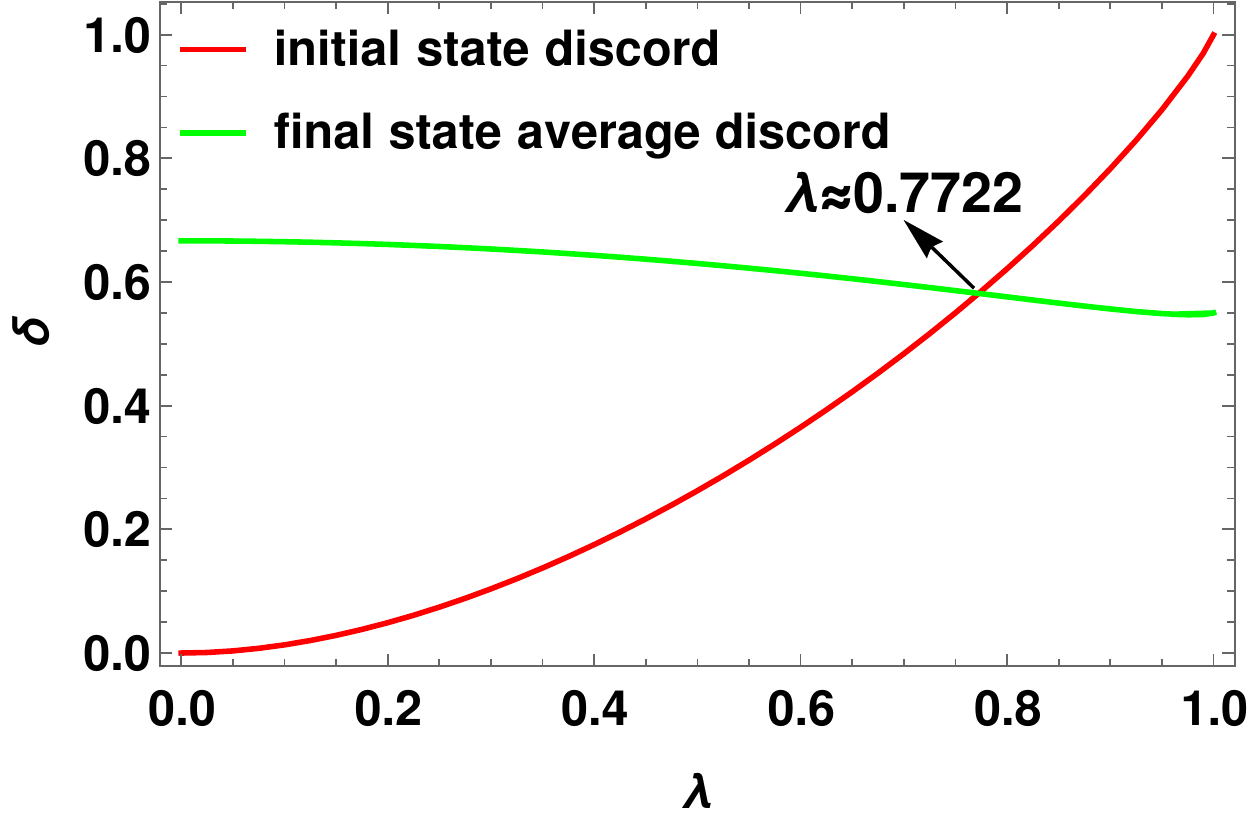}
  \caption{The red line shows the quantum discord as function of parameter $\lambda$ for the initial Alice's Werner state. The green line shows the average discord of the final states with variation in $\lambda$ (using the W state (\ref{shared state 3}) as the channel).}
    \label{fig5:}
\end{figure}

We can get similar kind of result if we take a 3-qubit state as a channel instead of the 4-qubit channel. However, Alice needs to send only 3 bits of classical information for this case.  Let's take a three qubit W state 
\begin{equation}\label{shared state 3}
\ket{\Psi}_{W} = \frac{1}{\sqrt{3}} \Big(\ket{0}_A\ket{01}_B + \ket{0}_A\ket{10}_B + \ket{1}_A\ket{00}_B\Big).
\end{equation}
We choose the measurement basis as 
\begin{eqnarray}\label{three dim basis}
&&\ket{b_1} = \frac{1}{2}\Big(\ket{000}+\ket{100}+\ket{011}+\ket{111}\Big),\nonumber\\
&&\ket{b_2} = \frac{1}{2}\Big(\ket{000}+\ket{100}-\ket{011}-\ket{111}\Big),\nonumber\\
&&\ket{b_3} = \frac{1}{2}\Big(\ket{000}-\ket{100}+\ket{011}-\ket{111}\Big),\nonumber\\
&&\ket{b_4} = \frac{1}{2}\Big(-\ket{000}+\ket{100}+\ket{011}-\ket{111}\Big),\nonumber\\
&&\ket{b_5} = \frac{1}{2}\Big(\ket{001}+\ket{010}+\ket{101}+\ket{110}\Big),\nonumber\\
&&\ket{b_6} = \frac{1}{2}\Big(\ket{001}-\ket{010}-\ket{101}+\ket{110}\Big),\nonumber\\
&&\ket{b_7} = \frac{1}{2}\Big(\ket{001}+\ket{010}-\ket{101}-\ket{110}\Big)\,\mbox{and}\nonumber\\
&&\ket{b_8} = \frac{1}{2}\Big(-\ket{001}+\ket{010}-\ket{101}+\ket{110}\Big).
\end{eqnarray}
We get similar kind of behavior like two previous examples. From Fig. \ref{fig5:}, it is clear that upto $\lambda\approx0.7722$ the average discord of final state is larger than the initial state.   

\begin{figure}
    \includegraphics[width=\linewidth]{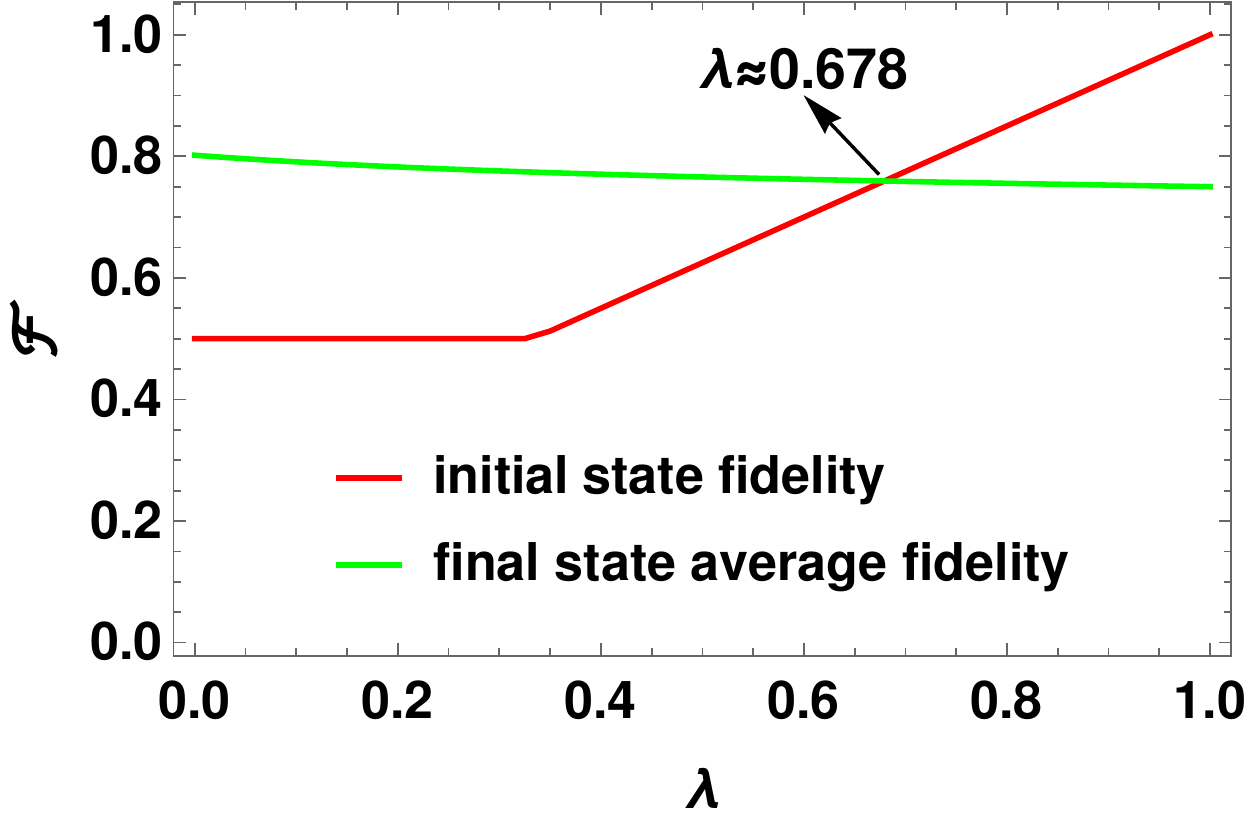}
   \caption{The red line shows the teleportation fidelity for the Werner state as variation in $\lambda$, the green line shows the upper bound or the maximum average fidelity possible for the final states res. as $\lambda$ is varied (using the cluster state given in \ref{sec:sec2.1} as the channel).}
   \label{fig2:}
\end{figure}

\section{Application}
\label{sec:sec3}

Essentially Werner state is maximally entangled state with some isotropic noise added. Consider a scenario where Alice wants to teleport a 1-qubit state using the Werner state as the channel. We ask the question that if the above protocol (see section~\ref{sec:sec2}) is applied and then the final states are used as channels instead of the initial Werner state, can one increase the teleportation fidelity?
Verstraete and Verschelde in \cite{verstraete} have found the upper and lower bounds of fidelity for any teleportation scheme. Given the channel $\rho$ the fidelity $\mathcal{F}^*(\rho)$ is  
\begin{equation}
\frac{1}{2}\bigg(1 + \frac{\mathcal{N}(\rho)}{1+\sqrt{1-(\frac{\mathcal{N}(\rho)}{\mathcal{C}(\rho)})^2}}\bigg) \leq \mathcal{F}^*(\rho) \leq \frac{1}{2}(1 + \mathcal{N}(\rho)), 
\end{equation}
where $\mathcal{N}(\rho)$ is the negativity \cite{Vidal} of the state $\rho$ and $\mathcal{C}(\rho)$ is the concurrence \cite{Wootters}. We calculate these two quantities as given in \cite{Vidal,Wootters} for the initial state and then find the bounds. For Werner state these two bounds are same. We do the same calculation for the final states (which can be obtained by applying our protocol as described in section \ref{sec:sec2}) to find the upper and lower bounds and then average it over. The average teleportation fidelity  $\overline{\mathcal{F}^*(\rho)}$ will belong to this range specified by lower and upper bound. We will compare our results graphically for those three examples described in the previous section. In the figures we have only showed the upper bound as this is the maximum achievable fidelity.

First we consider the example in subsection \ref{sec:sec2.1}. We find (as shown in Fig. \ref{fig2:}) that when the final states are used as the resource state instead of the Werner state, the upper bound or the maximum average teleportation fidelity is more than the fidelity of the initial Werner state (for most of the range of $\lambda$). Interestingly, we find that the upper bound of the average fidelity and discord both are decreasing with $\lambda$. This can be easily checked by looking at Fig.~\ref{fig1:} and Fig.~\ref{fig2:}. Moreover, we find that the $\lambda$ for which the average quantum discord of the final states is equal to the initial Werner state is different from the $\lambda$ where the teleportation fidelity using the Werner state is equal to the fidelity when using the final states. In this region one can see that average quantum discord for the final states is more, but the teleportation fidelity is less than the initial Werner state. From Fig.~\ref{fig1:} and Fig.~\ref{fig2:}, this region corresponds to the values from $\lambda \approx 0.678$ to $\lambda \approx 0.82047$. This difference may be due to the fact that quantum discord and quantum teleportation are not comparable. Proper optimization of the measurement basis may reduce this gap. Nonetheless, one important fact is that even when the initial Werner state is separable, we can use the final state as a channel for quantum teleportation. 

\begin{figure}[t!]
  \includegraphics[width=\linewidth]{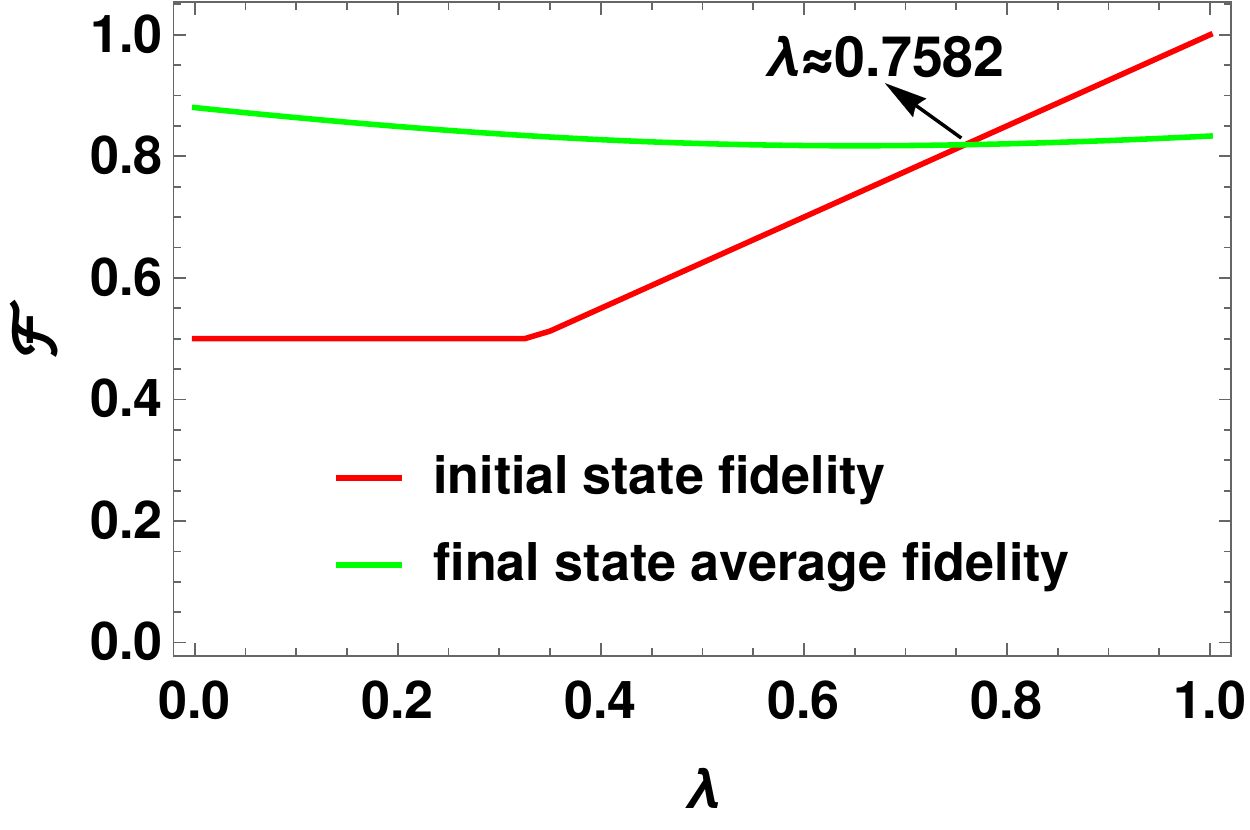}
  \caption{The red line shows the teleportation fidelity for the Werner state as variation in $\lambda$, the green line shows the upper bound or the maximum average fidelity possible for the final states res. as $\lambda$ is varied (using the state given in \ref{sec:sec2.2} as the channel).}
  \label{fig4:}
\end{figure}

Now we find the teleportation fidelity of the final state for the example given in  subsection \ref{sec:sec2.2}. We see a similar nature as the 4-qubit cluster state. In this example as we can see from Fig~\ref{fig3:} and Fig~\ref{fig4:}, the region of $\lambda$ (for which quantum discord for the final states are more in this region, but the teleportation fidelity is less than the initial Werner state) is from $\lambda \approx 0.7582$ to $\lambda \approx 0.8876$. 

\begin{figure}[t!]
  \centering
  \includegraphics[width=\linewidth]{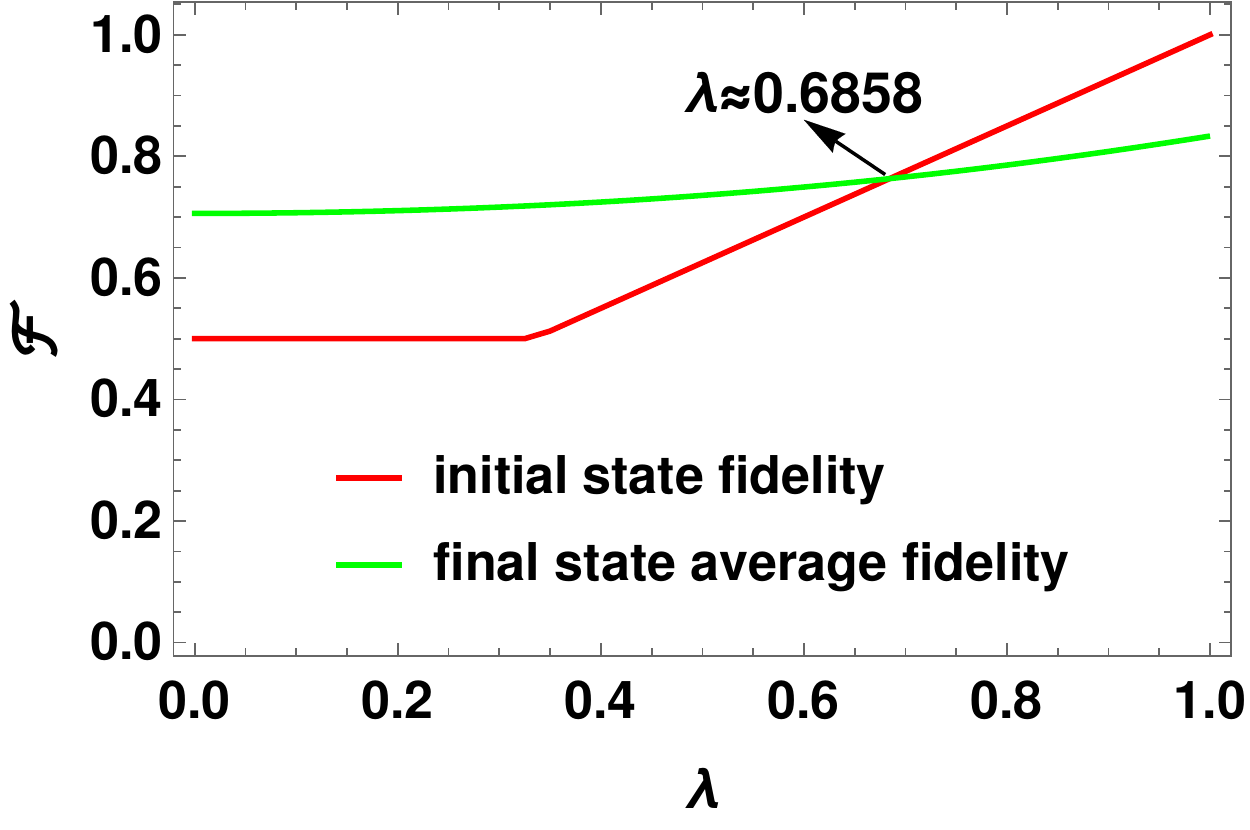}
  \caption{The red line shows the teleportation fidelity for the Werner state as variation in $\lambda$, the green line shows the upper bound or the maximum average fidelity possible for the final states res. as $\lambda$ is varied (using the state given in \ref{sec:sec2.3} as the channel).}
  \label{fig6:}
\end{figure}

Again we find the teleportation fidelity using the 3-qubit W state as the resource state. The measurement basis is given by Eq.~(\ref{three dim basis}). For the 3-qubit state the nature of teleportation fidelity is different from the 4-qubit states. From Fig.~\ref{fig5:} and Fig.~\ref{fig6:} we see that the teleportation fidelity behaves completely opposite of the quantum discord. The quantum discord decreases, but the teleportation fidelity increases as $\lambda$ increases. However, in this case the region (for which quantum discord for the final states are more in this region, but the teleportation fidelity is less than the initial Werner state) is from $\lambda \approx 0.6858$ to $\lambda \approx 0.7722$. Although we see for the examples considered above that when quantum discord is more than the initial state discord, the average teleportation fidelity is also more than the fidelity of the initial state (for some range of $\lambda$) but a direct connection can't be established from these results. These above examples give us some insight about the usefulness of discord in such a protocol. But whether it will be true for any general case needs to be further explored.

\section{Conclusion}
\label{sec:sec4}
We have shown that it is possible for Alice to prepare a state at Bob with a higher amount of quantum discord by sharing an entangled channel, local operations and classical communications. We also see that when the quantum discord of Alice's state increases then the average quantum discord of Bob's state decreases. Moreover, we showed above that the increment of the quantum discord in our protocol may be used as a way to increase fidelity of one qubit teleportation. Our results could also be applied to other quantum communication tasks like remote state preparation, entanglement distribution, etc. However our results are not optimized, so it might be possible that there could exist some basis for the given quantum channel such that the quantum discord of Alice's Werner state is always less than Bob's prepared state.

\section{Acknowledgment}
We would like to thank Satyabrata Adhikari and Prasanta Kumar Panigrahi for useful discussions.

\end{document}